\documentclass[]{elsart}

\usepackage{epsfig}
\usepackage{natbib}

\begin{document}

\begin{frontmatter}

\title{The RASNIK Real-Time Relative Alignment Monitor for the CDF Inner
Tracking Detectors}

\author[ucla]{David Goldstein},
\author[ucla]{David Saltzberg}

\address[ucla]{Dept. of Physics and Astronomy,
UCLA, Los Angeles, CA 90095-1547}


\begin{abstract}

We describe the design and operation of the RASNIK optical relative alignment system designed for 
and installed on the CDF inner tracking detectors.  The system provides low-cost minute-by-minute 
alignment monitoring with submicron precision.  To reduce ambiguities, we modified the original 
three-element rasnik design to a two-element one.  Since the RASNIKs are located within 10--40~cm 
of the beamline, the systems were built from low-mass and radiation-hard components and are 
operated in a mode which reduces damage from radiation.  We describe the data-acquisition system, 
which has been running without interruption since before the CDF detector was rolled into its 
collision hall in March 2001.  We evaluate what has been learned about sources of detector motion 
from almost two years of RASNIK data and discuss possible improvements to the system.

\end{abstract}
~\vspace*{-0.5in}\\

\end{frontmatter}


\newpage

\section{Introduction}

A critical feature of the major upgrade of the CDF experiment~\cite{cdftdr} at Fermilab is its 
inner silicon trackers.   These detectors provide the ability to reconstruct and trigger on 
clusters of charged tracks originating $100$ to $1000$~$\mu$m away from the primary interaction.  
Such detached vertices of charged particles, which may indicate the  presence of a particle 
containing a $b$ quark, are critical to the CDF physics goals: characterization of top quark 
production and decay; measurements of heavy flavor mixing; CP violation and rare particle decays; 
searches for supersymmetric particles and, eventually, searches for the Higgs boson.  However, 
the ability to resolve the separation between the detached vertex and primary interaction 
requires that the silicon trackers remain stable to approximately $10$~$\mu$m.

While detector alignment can be accomplished with tracks taken over a long integration time, the 
RASNIK technique (Relative Alignment System of the Nikhef Institute)~\cite{rasnikwww} offers a 
low-cost tool to observe motion on the time scale of seconds with sensitivity better than one 
micron.  RASNIK systems are being developed for the Atlas~\cite{atlasrasnik} and 
CMS~\cite{cmsrasnik,cmsrasnik2} experiments which are currently being installed at the Large 
Hadron Collider at CERN.  These systems have already been installed in a running experiment---the 
CHORUS experiment at CERN used RASNIKs~\cite{chorusrasnik} to monitor the alignment of detectors 
used to track particles back into a photographic emulsion.

In section~\ref{design} we describe the RASNIK principle and the modifications we made to 
specifically adapt it to CDF's silicon inner tracker.   In sections~\ref{implementation} 
and~\ref{data} we describe the specific implementation at CDF and discuss the data.  In 
section~\ref{improvements} we discuss possible improvements to the system.

\section{The Two-Element RASNIK}
\label{design}

\begin{figure}
\begin{center}
\leavevmode
\epsfxsize=4.0in
\epsfbox{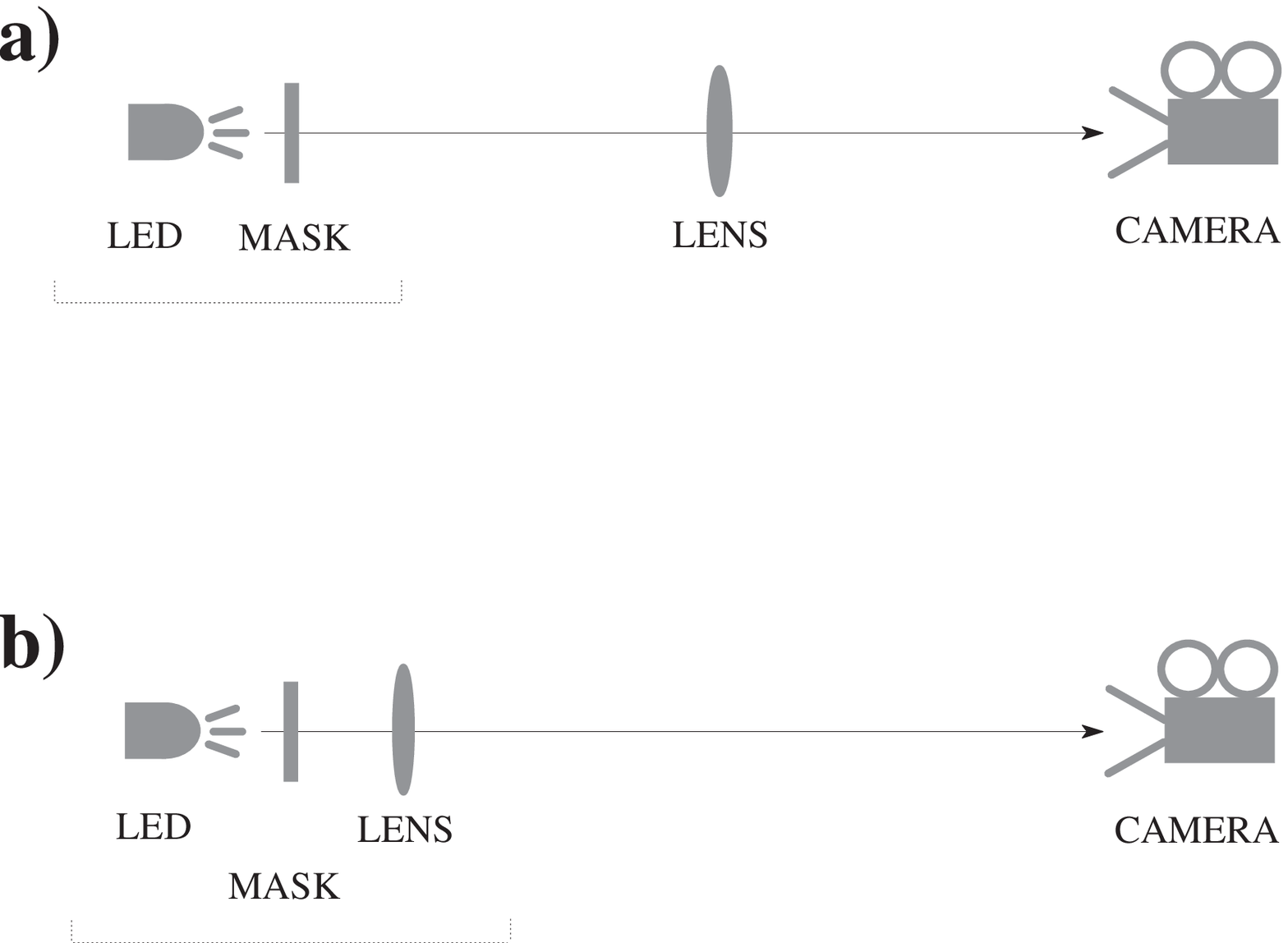}
\vspace{1in}
\caption{
The basic features of the RASNIK technique for the a) original three-element
system and b) the two-element system developed for CDF.  Incorporating the lens
into the projector piece introduces system-specific magnification; this complicates 
position analysis, but eases installation and reduces ambiguity.  Without this 
modification the installation of RASNIKs in the environment at CDF would not 
have been feasible.
}
\label{rasprinc}
\end{center}
\end{figure}

\begin{figure}
\begin{center}
\leavevmode
\epsfxsize=4.0in
\epsfbox{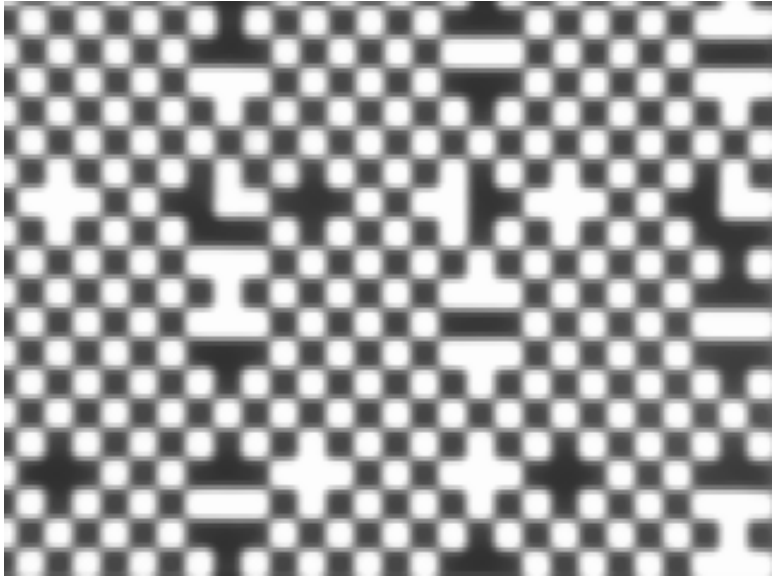}
~\vspace*{.5in}
\caption{
A typical image from the CMOS camera after capture by the frame grabber.
Each small square is 20~$\mu$m on each edge.  The interrupts which encode the
global position on the mask are visible as irregularities in the checkerboard pattern.
}
\label{mask}
\end{center}
\end{figure}

The original RASNIK principle is based on projecting the image of a finely detailed mask through 
a lens onto a small digital camera, as shown in figure~\ref{rasprinc}a.  The three components 
(mask, lens, camera) are mounted on separate detector subsystems and/or support structures.  
If any one of these three elements moves there will be a corresponding movement of the projected 
image on the camera.  In this fashion, the system tracks the relative alignment of the pieces.  
A typical mask in such a system is a grid of black and white squares of order 100~$\mu$m size, 
photo-etched onto a thin quartz slide.  As shown in figure~\ref{mask}, deviations from a perfect 
grid are coded into the pattern to remove large scale position ambiguities; it is therefore 
called a ``coded mask''.  The use of such a pattern provides great statistical power since there 
are hundreds of black/white transitions to measure in a given image.

\begin{figure}
\begin{center}
\leavevmode
\epsfxsize=5.0in
\epsfbox{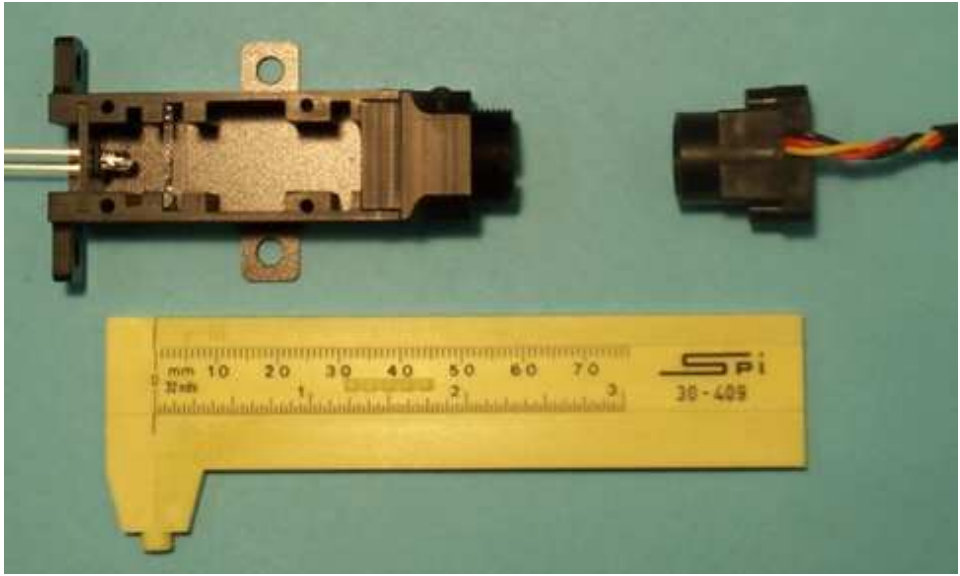}
~\vspace*{.5in}
\caption{
A photograph of the RASNIK projector (upper lid removed) and camera 
used in the CDF RASNIK system.  The LED and coded mask are visible
towards the left (back) of the projector interior.  The lens is 
located within the threaded barrel at the front of the piece.  Note 
that the distance of the mask to the lens may be changed to accommodate 
different baselines.  The lens position may be determined by final 
focussing {\it in situ}.
}
\label{rasphoto}
\end{center}
\end{figure}

For our application on the CDF trackers we removed one movement ambiguity by integrating the light 
source, coded mask and lens into a single ``projector'' piece.  The asymmetry of the technique 
introduces magnification, so new masks using smaller ($20.00\pm0.01)$~$\mu$m squares were printed. 
The resultant two-element system is compared to the original three-element technique in 
figure~\ref{rasprinc}b.  Because the mask is coded, at least 10 squares (contiguously in both 
horizontal and vertical directions) must be visible to the camera in order to reconstruct the 
global position.  Since the magnification changes for different projector-camera distances, 
``baselines'', we provided the projector with several slots for positioning the coded mask.  
Hence the magnification can be kept roughly constant, independent of the particular baseline, by 
a small change in the placement of the coded mask.  Providing multiple locations for the mask 
also eliminates the need for a diffuser, allowing the use of a single LED and further reducing 
the power consumption and mass of the piece.  Since the depth of focus is typically of order 
1~mm, the lens is mounted in a threaded barrel so that final focus can be achieved after both the 
projector and camera are glued in place.  Figure~\ref{rasphoto} shows a photograph of the actual 
projector and camera pieces used in CDF.

Because the elements are placed so close to the beamline they have to be low-mass and 
radiation-hard.  The projector piece is made of thin polycarbonate, which provides ample 
rigidity for components of this small size.  A projector {\it in situ} (with mask, lens, and LED 
installed) weighs approximately 20 grams.  A typical particle trajectory which traverses the 
piece corresponds to approximately 1\% of a radiation length.  Various optical materials were 
tested in the UC Davis cyclotron, exposing them to $\sim$200~kRad using a 63~MeV proton 
beam~\cite{radtesting}.  This roughly corresponds to the maximum integrated dose expected for an 
inner (SVX) RASNIK over the course of CDF's Run IIa.  As a result of these tests we determined 
that a standard infrared LED\footnote{In general, LED's which were described by the manufacturer 
as ``rad hard'' did not necessarily perform any better than more randomly chosen units.  Infrared 
emitters were chosen to match the peak sensitivity of the camera.} and BK7 glass lenses were 
adequately radiation hard, and used these components in all systems installed on the detectors.  
We replaced the CCD camera typically used in previous RASNIK systems with a CMOS-based camera.  
Such cameras are smaller and are known to be more robust to radiation.  Still, we observed that 
the camera would suffer fatal damage at this dose, but only if it was powered on during the 
exposure.  No visible damage was apparent if the camera was turned off.  Since the systems are 
read out one at a time, we energize each camera (and LED) only when it is being used, and expect 
no significant radiation damage.  This mode of operation corresponds to an approximate 5\% duty 
cycle.  After 1.5 years of running, including higher than expected beam losses, no visible 
effects of radiation damage have been observed on any system.  

\begin{figure*}
\leavevmode
\hspace*{-.3in'}
\epsfxsize=3.0in
\epsfbox{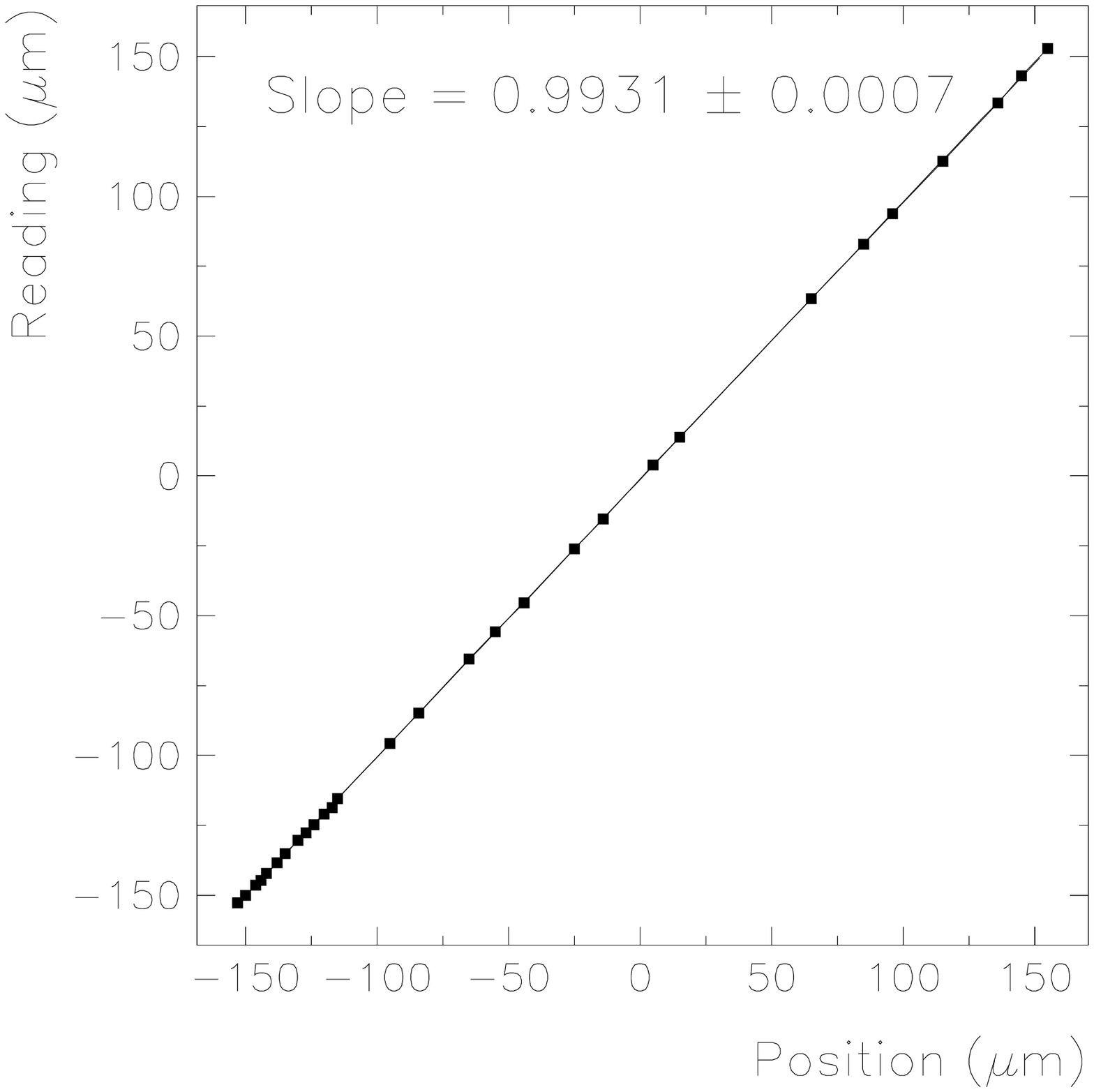}
\epsfxsize=3.0in
\epsfbox{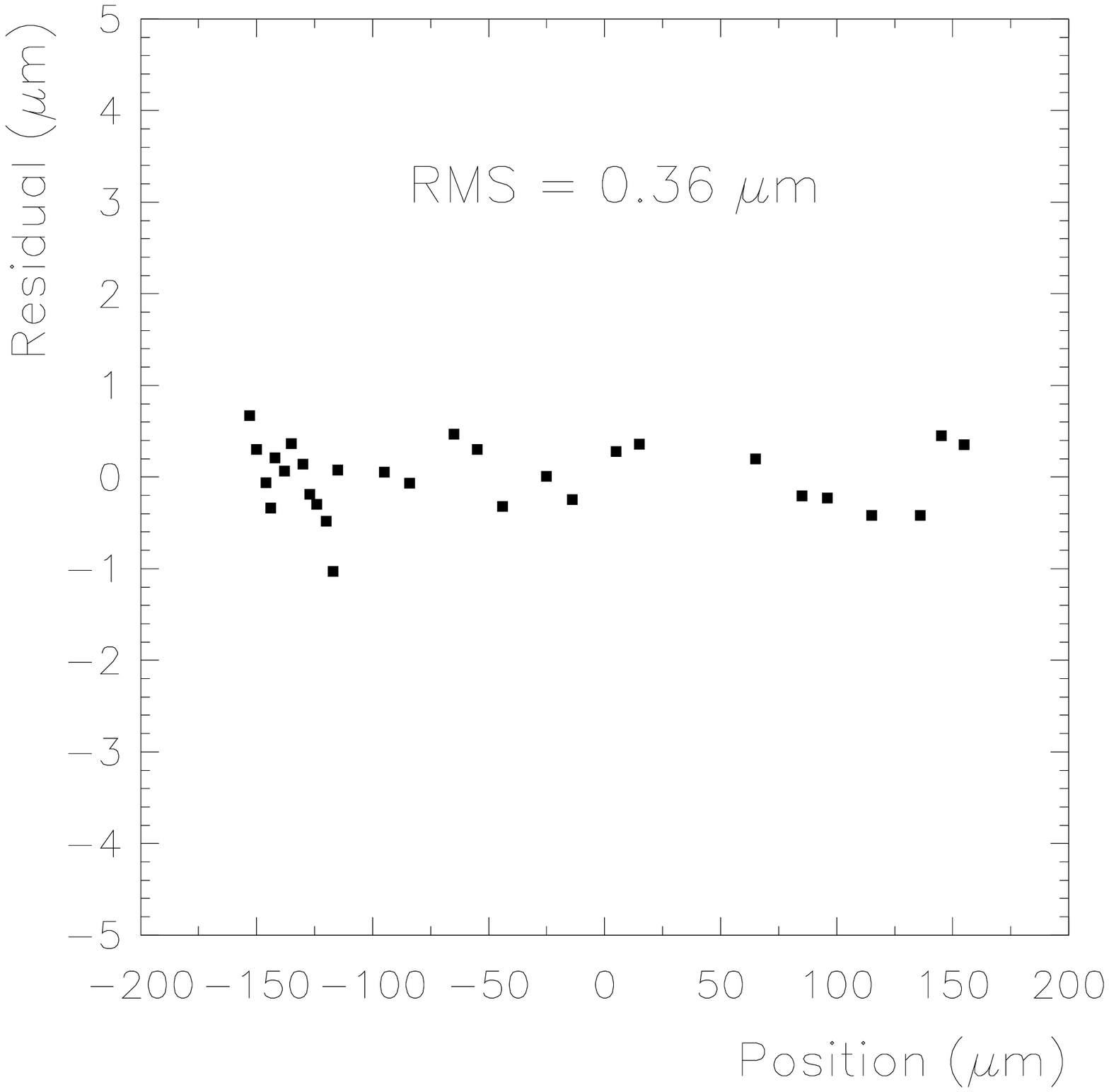}
~\vspace*{.5in}
\caption{
Testbench results of the two-element RASNIK.  \underline{Left}: Reconstructed
vs. actual camera position. \underline{Right}: Residuals.   Note that the
0.4$\mu$m residuals probably reflect our ability to position the moveable stage, 
rather than the limiting resolution of the RASNIK.
}
\label{testing}
\end{figure*}

The accuracy of the two-element system was tested by placing the projector on a moveable stage on 
an optical bench.  Figure~\ref{testing} shows the actual vs. reconstructed position of the camera 
as it was moved horizontally.  Similar results were obtained for vertical movement.  The figure 
also shows the residuals, which are unbiased with an RMS of 0.4~$\mu$m.  Such accuracy is at the 
limit of our ability to position the camera and probably does not indicate the limiting RASNIK 
resolution.

\section{CDF Implementation \& Data Acquisition}
\label{implementation}

\begin{figure}[ht]
\begin{center}
\leavevmode
\vspace*{-2.5in}
\epsfxsize=6.5in
\hspace*{-.5in}
\epsfbox{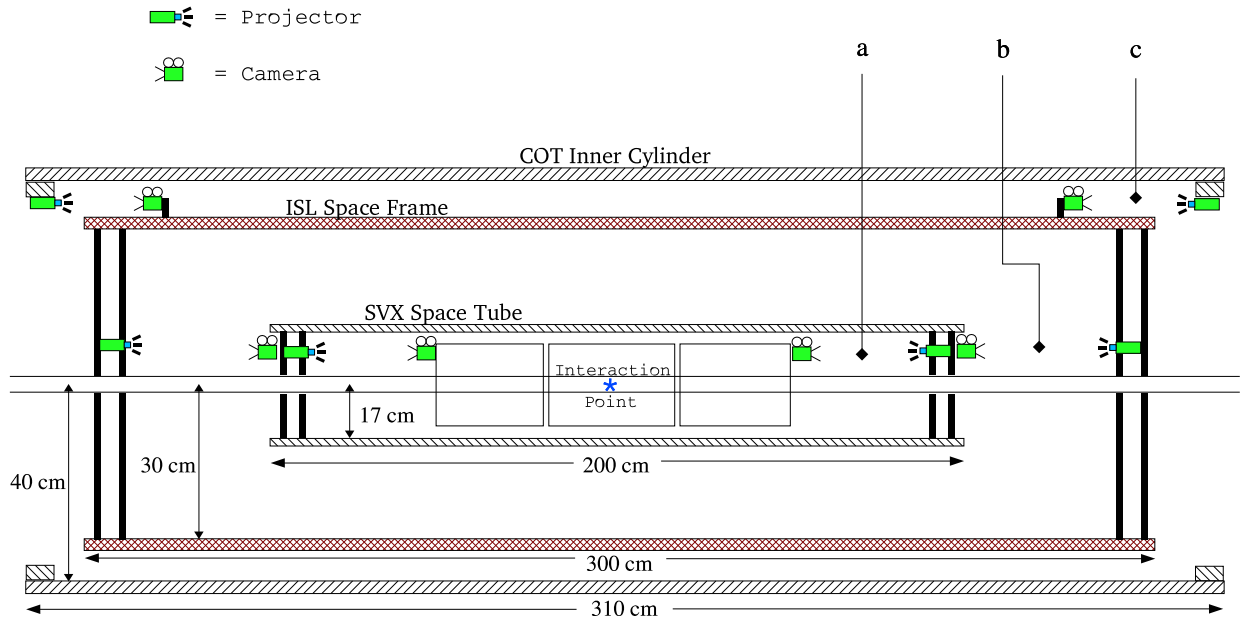}
~\vspace*{-3.0in}
\caption{
Diagram (not to scale) showing the placement of representative RASNIK systems inside
the silicon tracking system.  a) SVX, b) SVX$\rightarrow$ISL, c) ISL$\rightarrow$COT 
(inchworm).  The mirror symmetry of the placement is shown here.   The systems are 
repeated every $120^{\circ}$ in azimuth (for clarity this is not shown here) yielding 
a total of 17 (There is one exception to the symmetry, described in the text, which 
results in 17 instead of 18) systems.
}
\label{placement}
\end{center}
\end{figure}

CDF's silicon system consists of three independent elements:  a single layer mounted directly to 
the beampipe known as Layer~00, an inner tracker (radii: $2.4-10.6$~cm) known as the SVX, and an 
outer tracker (radii: $20.0-28.0$~cm) known as the ISL.  The entire superstructure is connected 
at its ends to the inner wall of CDF's drift chamber (central outer tracker, COT) through 
piezoelectric positioners, called ``inchworms''~\cite{inchworms}.  Seventeen RASNIKs are 
strategically placed\footnote{The inchworm system employs a gimbal mount at one of its six 
support points on which it is not possible to mount a rasnik.  This breaks the placement symmetry 
described in this paragraph and implies one fewer installed system.} throughout the silicon 
tracker system so that movements of particular systems can be combined to determine the global 
twisting or translational motion of the silicon superstructure.  The placement follows a 
three-fold azimuthal symmetry (following that of the inchworms) with mirror symmetry about the 
interaction point.  A schematic diagram of the placement of the RASNIK systems in CDF is shown in 
figure~\ref{placement}.  In addition to the 17 deployed RASNIKs, two systems (one short and one 
long-baseline) are installed as a control in the DAQ rackspace.  These are mounted on 1/2-inch 
aluminum plates which are set on vibration damping material, so little relative motion is 
expected.

The data-acquisition is controlled by a 600~MHz rack-mounted PC running Linux, located in CDF's 
counting room.  The data-taking proceeds rasnik-by-rasnik through a simple endless loop.  For 
each RASNIK, the PC activates two relays corresponding to low-voltage power (for both the LED and 
camera) and the output video signal.  All signals are carried on $\sim$200 ft. lengths of 
shielded twisted pair cable; no degradation of the video signal has been observed.  Care was 
taken to avoid ground loops.  When the relay is closed the video signal is connected to a PCI 
``frame grabber'' card which takes the picture.  The image is reconstructed in roughly a second, 
yielding transverse positions (in the plane of the camera's image sensor), a rotation angle of 
the mask, and several quality-of-fit variables.  In addition, a crude ($\sim 100\mu$m) measurement
of the longitudinal distance between the camera and projector is obtained from the change in 
magnification of the image.  In standard operating mode each RASNIK's position is measured 
approximately every 120 seconds.  The minimum time necessary to take a data point is limited by 
the camera:  Since the systems are normally unpowered to mitigate damage from radiation, a few 
seconds must be allowed upon power-up before the camera output signal becomes stable.  For 
special running conditions, such as when the silicon detector position is being changed, a single 
RASNIK's position can be read out every second.  The readout system has proven highly reliable; 
data-taking has proceeded without interruption for almost two years.  The raw data is archived on 
disk, and plots of RASNIK data from the current and previous month are available on a web page 
which is automatically updated every hour.  The direct video output from the systems is also 
viewable in real-time on a monitor channel in the CDF control room.  A very useful feature, which 
we plan to implement soon, would be to add a bit to the raw RASNIK data which signifies whether 
physics data was being taken with the CDF detector.  The existing plots on the web page will then 
display these periods in a different color, so that they will be easily recognizable at a glance.

It is worth noting that the RASNIKs are quite adaptable, and can also be of use as a stand-alone 
setup, without collecting any data.  In October 2000, the silicon detector mounted directly onto 
the beampipe (Layer~00) was inserted into the inner silicon tracker (SVX).  This was a delicate 
process, with very little clearance margin.  Two RASNIKs were orthogonally mounted downstream of 
the insertion point, with cameras attached to the beampipe and projectors mounted close by on 
makeshift stands.  The RASNIK images were displayed continuously on two adjacent monitors, and 
provided an instantaneous visual cue to any incidental contact during the insertion.

\section{Data}
\label{data}

\begin{figure}
\begin{center}
\leavevmode
\epsfxsize=4.0in
\epsfbox{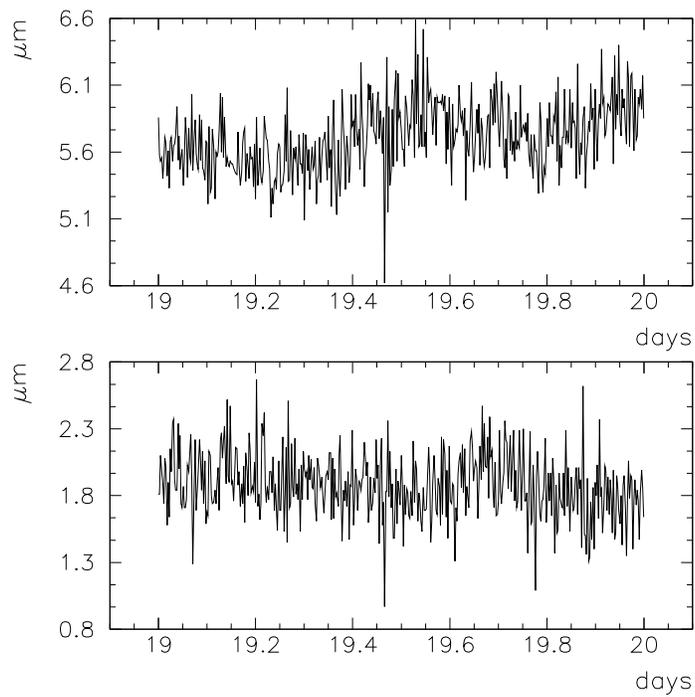}
~\vspace*{.5in}
\caption{
 Data taken from a ``control'' RASNIK system over a 24~hour period.  
Transverse motions ($x$ and $y$) are displayed in the upper and lower
plots respectively.  This RASNIK was specially mounted and 
isolated from vibrations, so little movement is expected.  The RMS of the 
movement is ~$<0.5\,\mu$m.
}
\label{rascontrol}
\end{center}
\end{figure}

As described in the previous section, two RASNIKs were installed in the DAQ rackspace as a control.
The data from one day of data-taking from one of these systems are shown in 
figure~\ref{rascontrol}.  The RMS of the apparent relative motion is $< 0.5$~$\mu$m; data from 
the other control system were similar.

\begin{figure*}
\leavevmode
\epsfxsize=3.0in
\epsfbox{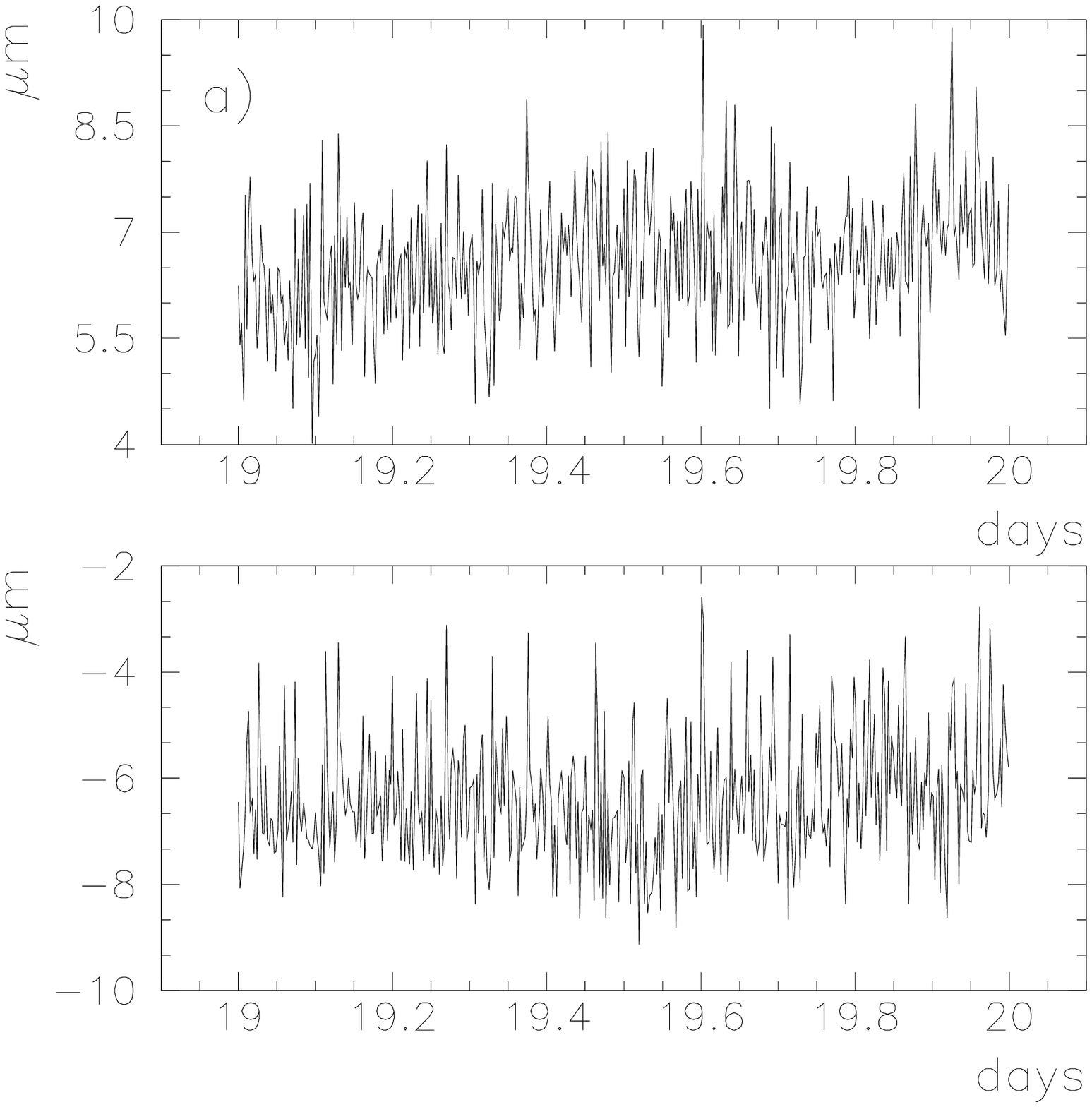}
\epsfxsize=3.0in
\epsfbox{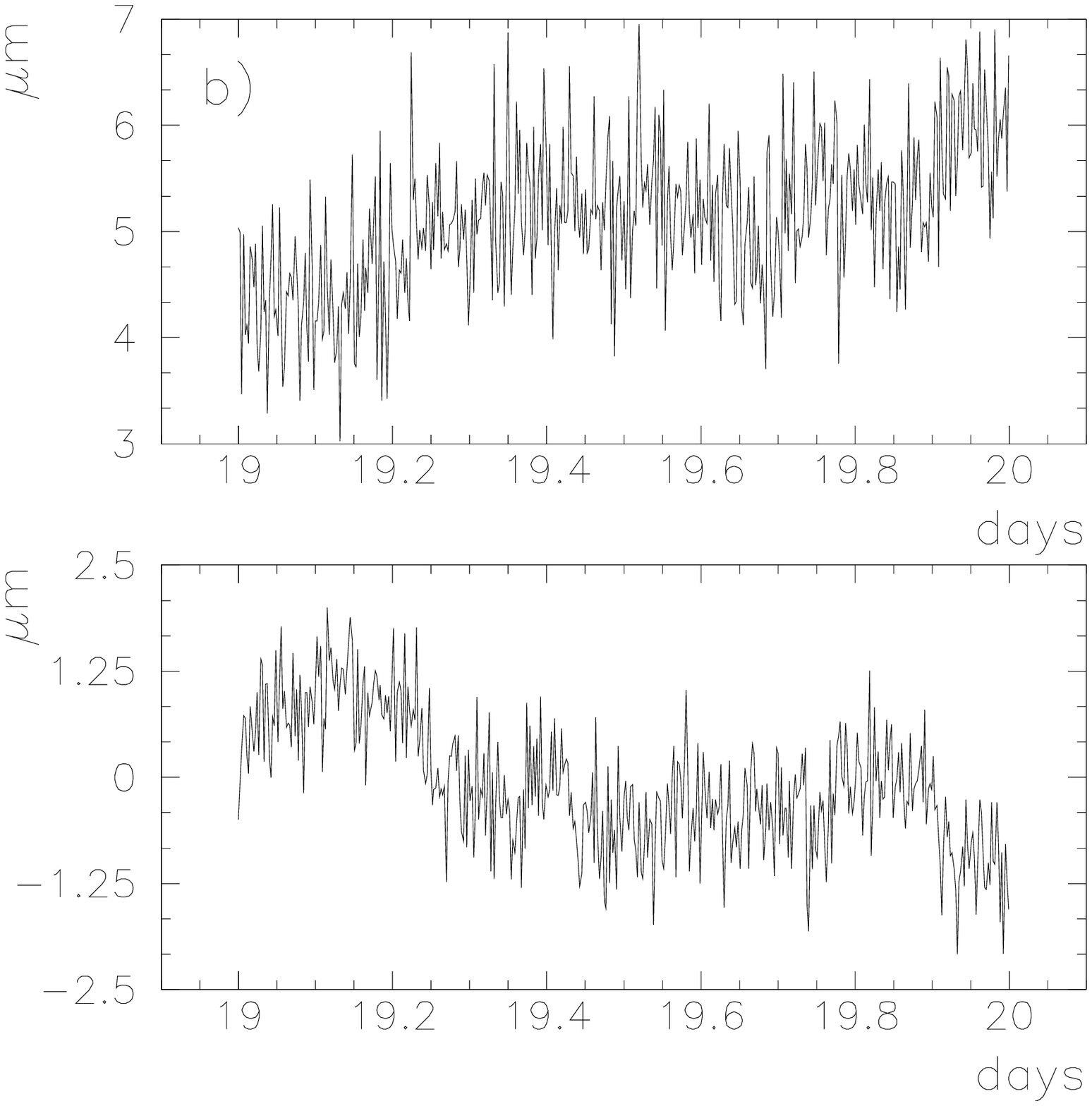}
\begin{center}
\epsfxsize=3.0in
\epsfbox{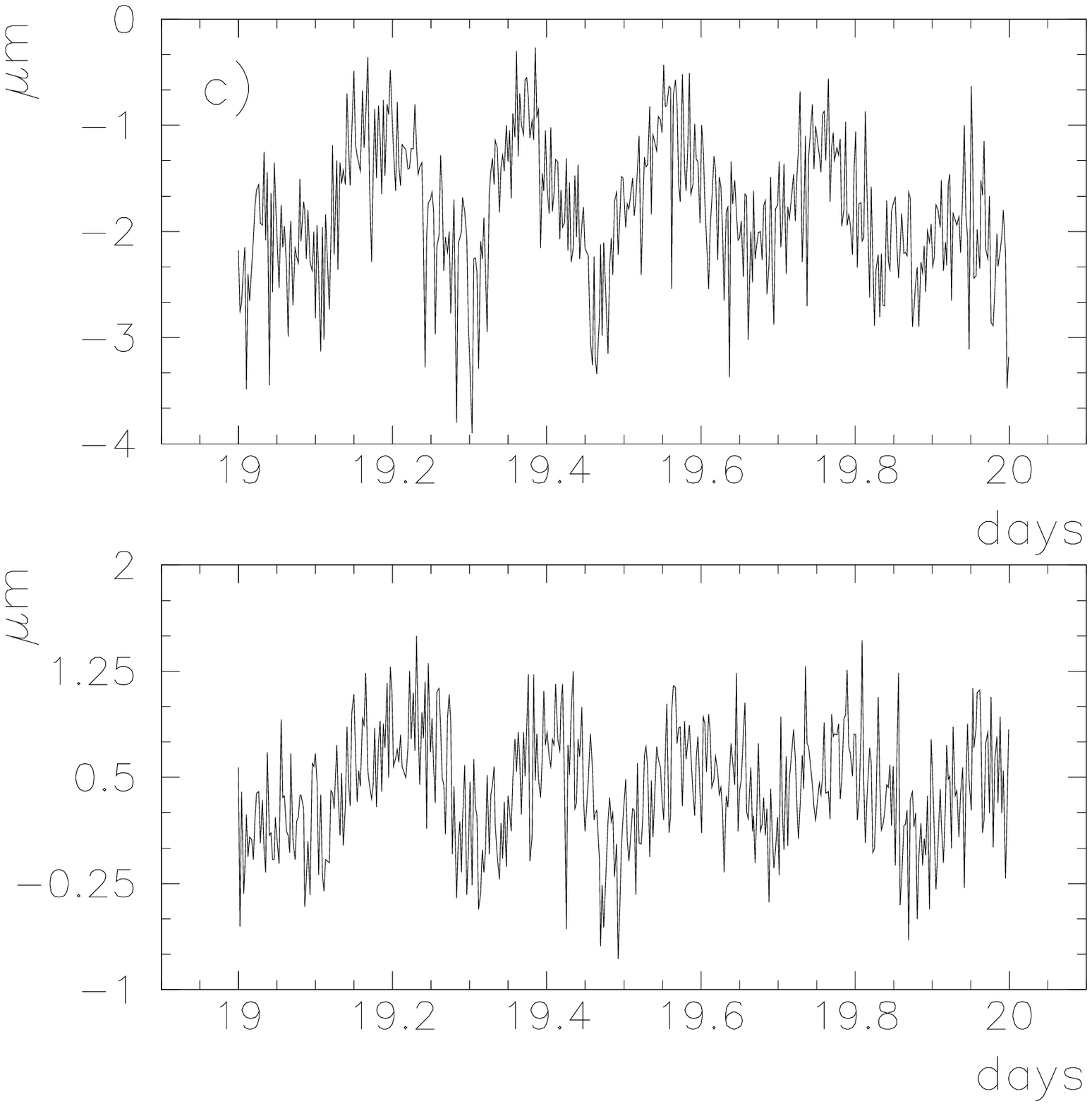}
~\vspace*{.5in}
\caption{
Transverse motion ($x$ and $y$) 
data taken from one of each of the three types of
RASNIK systems over a particularly stable 24~hour period, {\it i.e.},
with no changes to the solenoid current or silicon power.   
a) SVX, b) SVX$\rightarrow$ISL,  c) ISL$\rightarrow$COT (inchworm).  The RMS of the 
movements during periods with these types of environmental conditions is 
typically a few microns.
}
\label{quietday}
\end{center}
\end{figure*}

Each of the 19 RASNIK systems takes data points in the two transverse ($x,y$) and longitudinal 
($z$) directions every couple of minutes.  In the following figures we show some representative 
features of the data taken thus far.  Figure~\ref{quietday} shows one day of motion recorded in 
the $x$ and $y$ directions for a quiet period, when CDF was taking data but no changes were made 
to any voltages or to the solenoid current.  The data are shown for one each of the SVX, 
SVX$\rightarrow$ISL and ISL$\rightarrow$COT (inchworm) systems.  The data reflect stability to 
$< 5.0$~$\mu$m.  This is a slightly higher variance than the control systems, but is within 
design tolerances for the CDF tracking detectors.

\begin{figure*}
\leavevmode
\epsfxsize=3.0in
\epsfbox{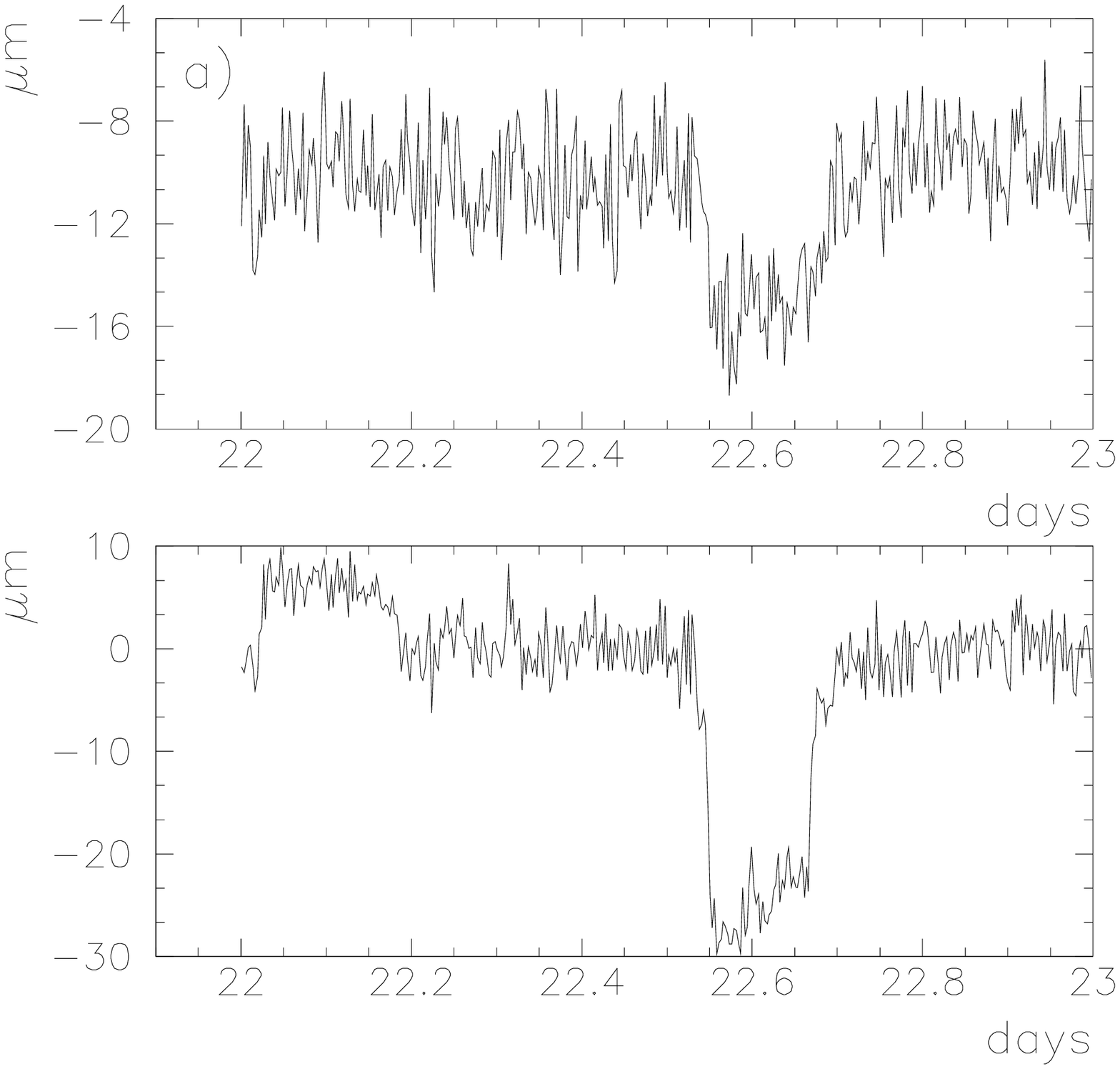}
\epsfxsize=3.0in
\epsfbox{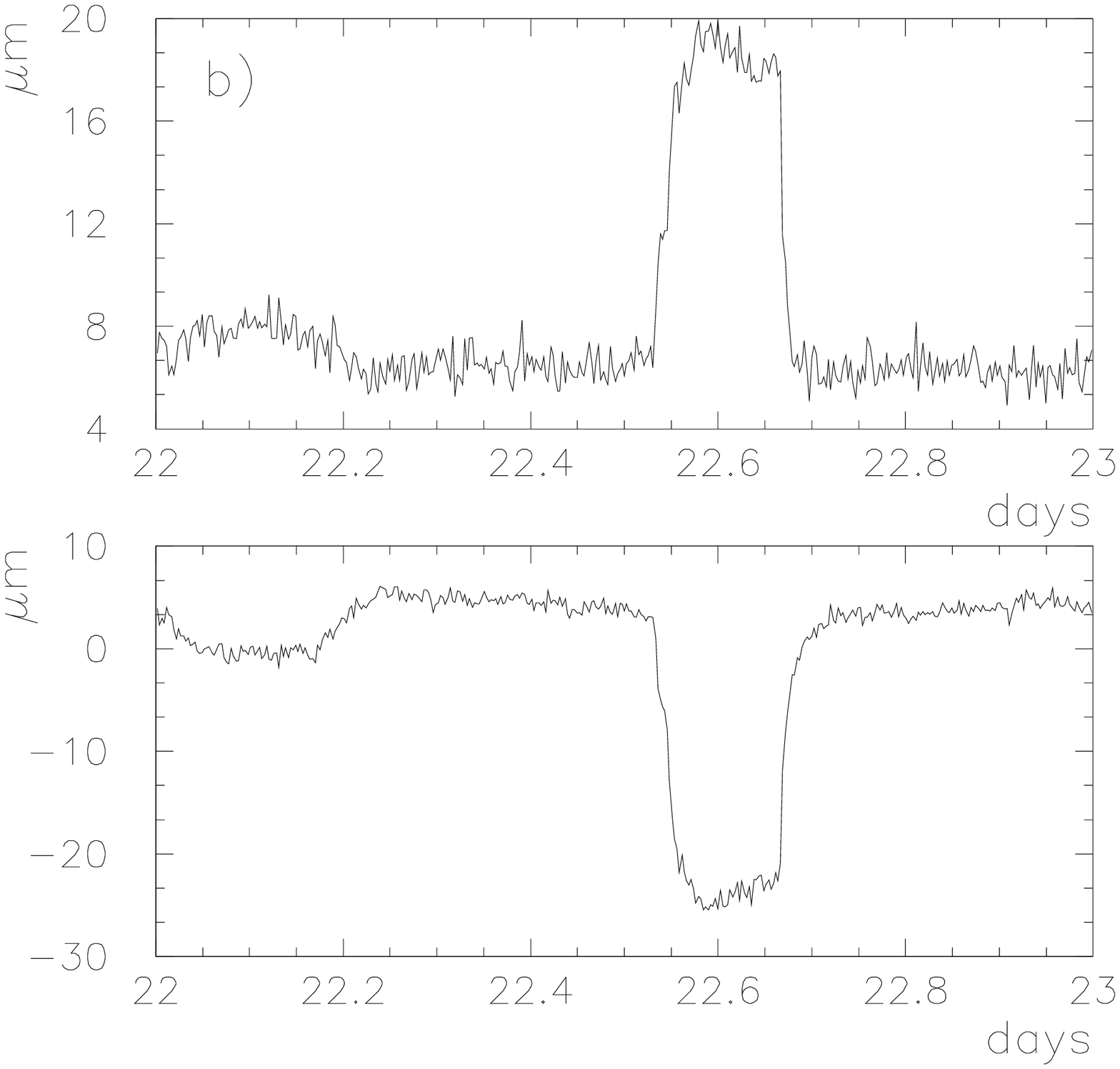}
\begin{center}
\epsfxsize=3.0in
\epsfbox{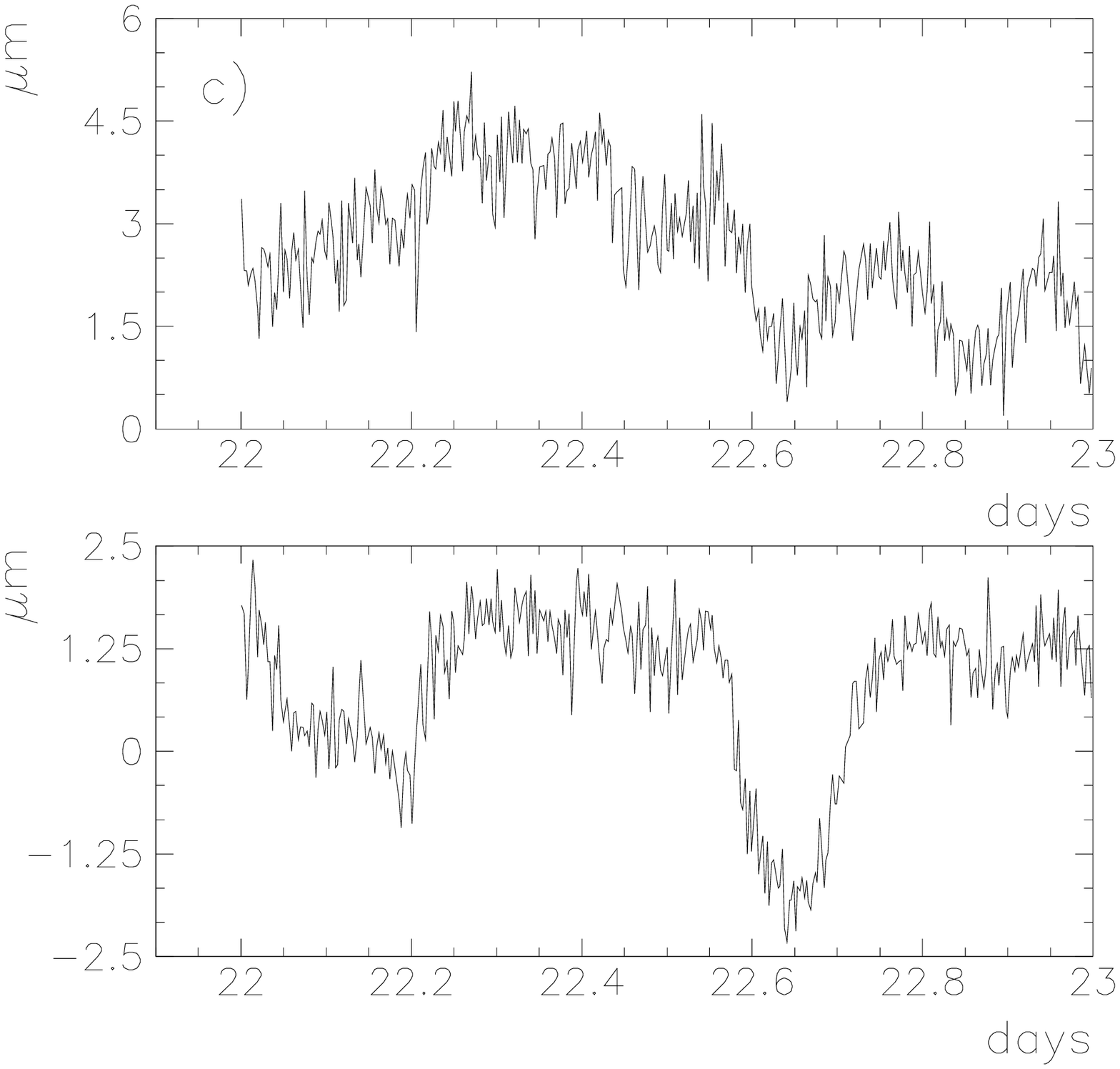}
~\vspace*{.5in}
\caption{
Transverse motion ($x$ and $y$) data taken from one of each of the three types of
RASNIK systems over a more typical 24~hour period.  a) SVX, b) SVX$\rightarrow$ISL,  
c) ISL$\rightarrow$COT (inchworm).  The large motion at 12:50 (22.55) corresponds to a change
in the solenoid current.  Note however that the silicon returns to its nominal 
position during the conditions of data-taking.
}
\label{typday}
\end{center}
\end{figure*}

\begin{figure*}
\leavevmode
\epsfxsize=3.0in
\epsfbox{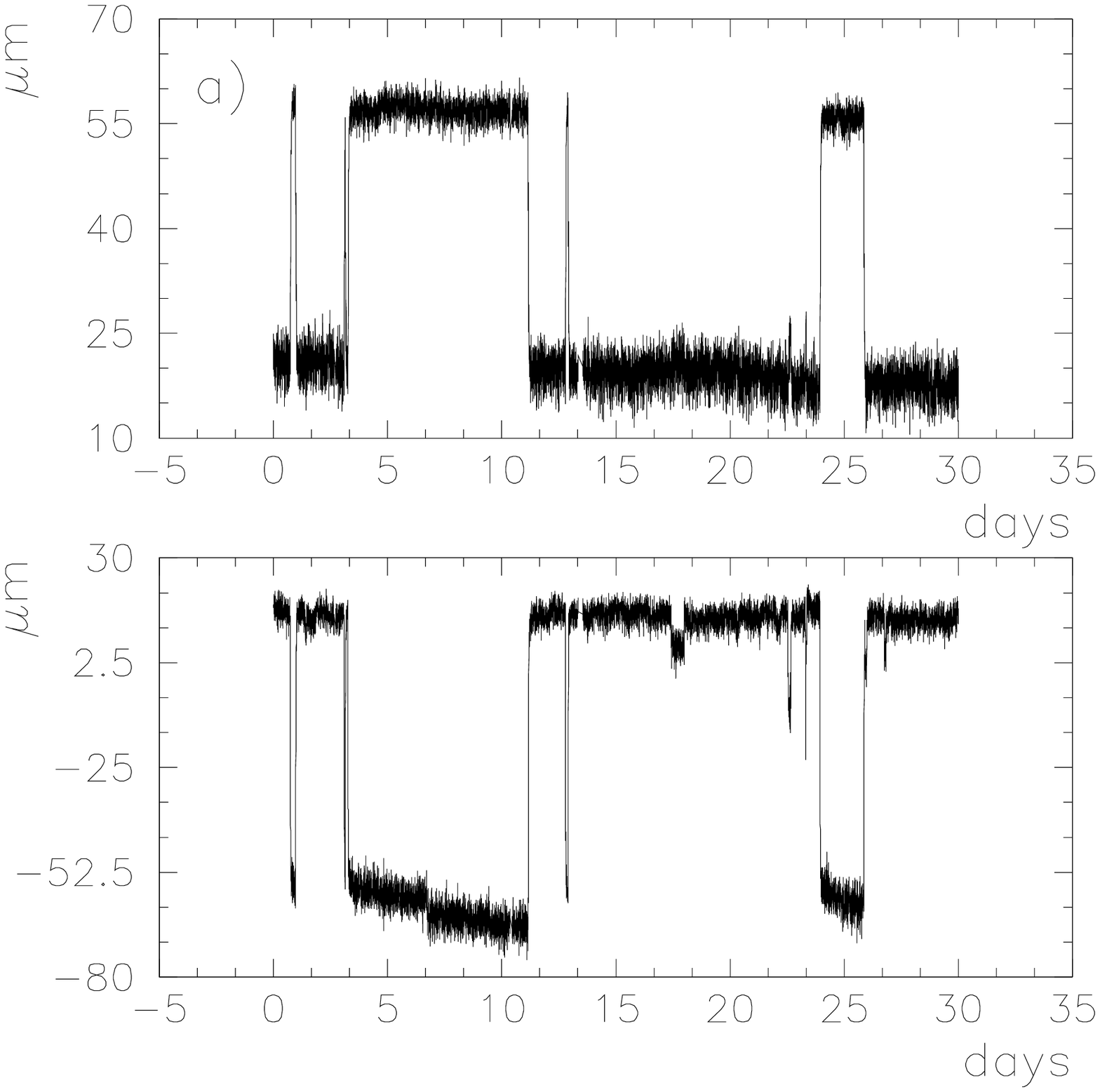}
\epsfxsize=3.0in
\epsfbox{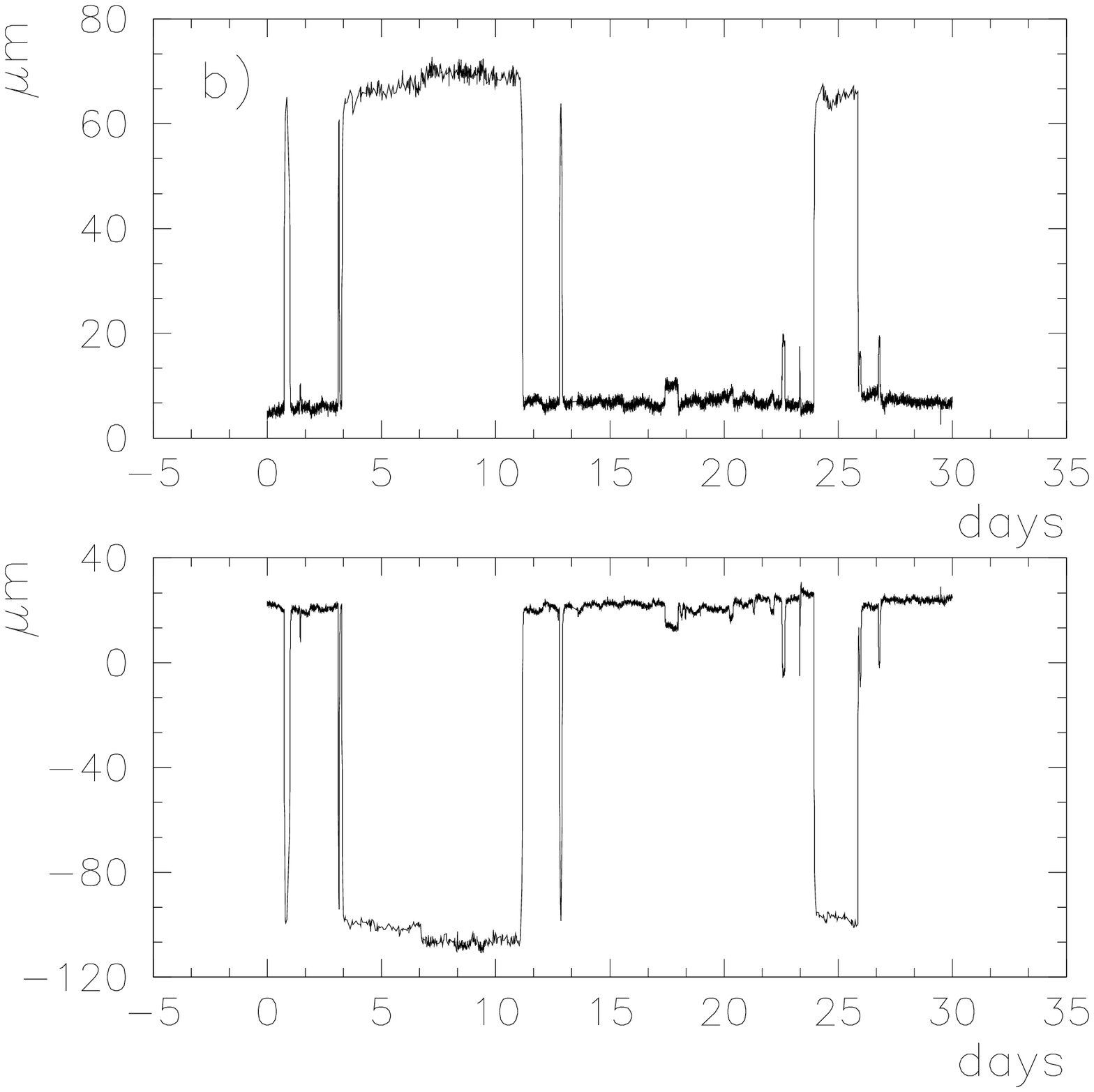}
\begin{center}
\epsfxsize=3.0in
\epsfbox{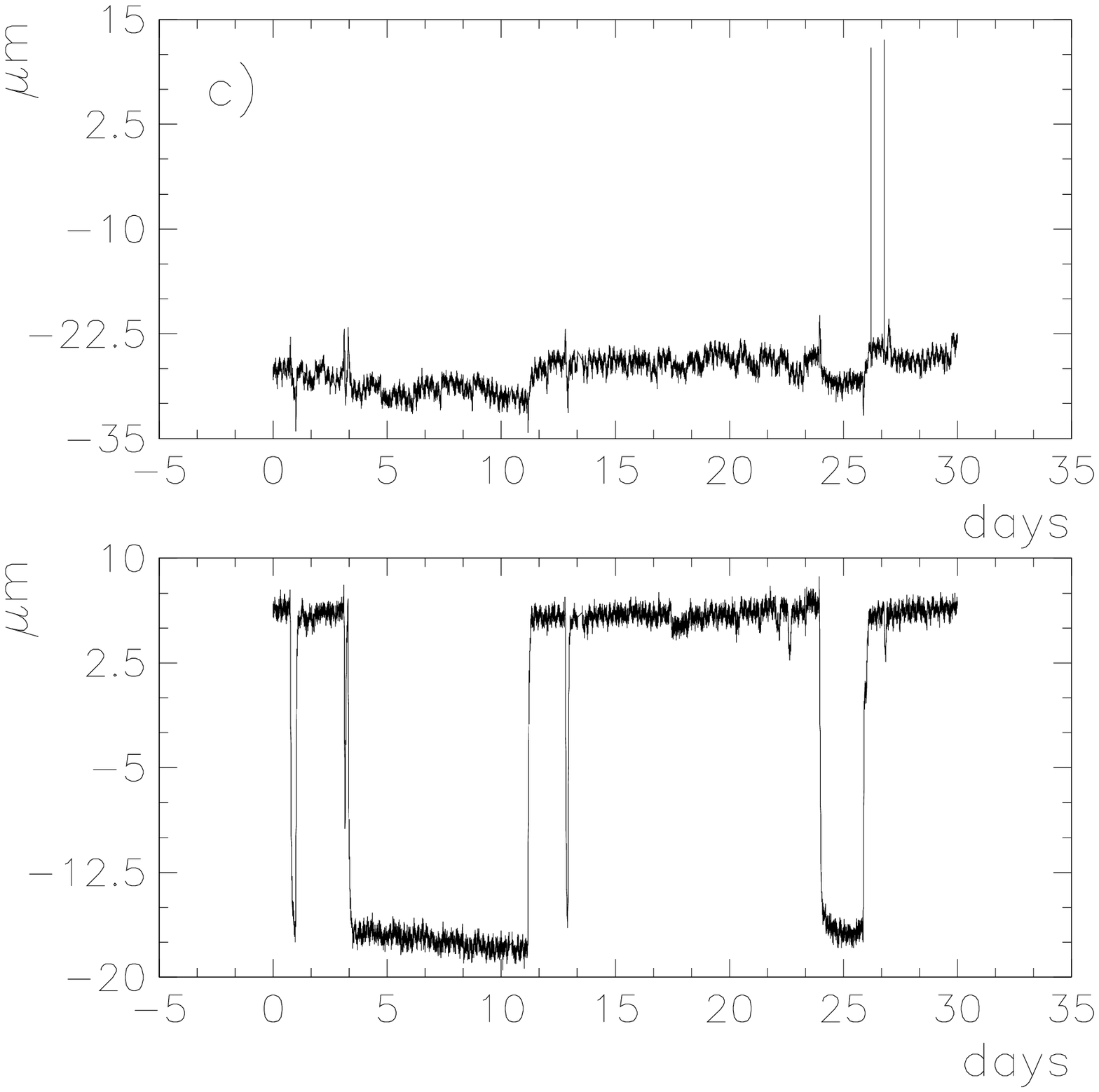}
~\vspace*{.5in}
\caption{
Transverse motion ($x$ and $y$) data taken from one of each of the three types of
RASNIK systems over a typical one-month period.  a) SVX, b) SVX$\rightarrow$ISL,  
c) ISL$\rightarrow$COT (inchworm).  Many of the large motions due to energizing the 
solenoid etc. are visible.
}
\label{typmonth}
\end{center}
\end{figure*}

A more typical day shows large motions, as shown in figure~\ref{typday}.  The $\sim 30$~$\mu$m 
feature at 12:50 (at 22.55 days) corresponds exactly to a change in the current in CDF's 1.4~T 
solenoid.  A view of an entire month is shown in figure~\ref{typmonth}.  Many of the large 
motions due to energizing the solenoid etc. are visible.  It is worth noting however that during 
data-taking periods with silicon included in the run, the entire system always returns towards 
its nominal position.  The overall behavior thus far in CDF's Run IIa seems consistent with some 
settling of the detector subsystems:  Over the course of a month, some creep in the nominal 
position has sometimes occurred.  This was more prevalent early on, and has tended to diminish 
with time.  Since the RASNIKs have shown that the silicon trackers are very stable during 
data-taking, the main long term use of the system will be to flag times when new global alignment 
periods should be defined. 

When considering large motions recorded by some of the systems, it is important to recognize that 
changes in the angle of the projector piece with respect to the camera can be interpreted as 
large linear motions as the projected image sweeps across the camera.  Because of the placement 
of the systems at CDF, we can disentangle these effects by performing a global analysis of the 
motion of all the RASNIKs together.  For example, when the cooling to the silicon detectors is 
switched on or off, a mechanical coupling between the beampipe and the inner SVX/ISL structural 
supports causes the SVX and ISL systems to record a linear relative translation which is 
approximately 10 times the actual movement.  With the existing system, this effect is discernible 
by a global analysis of the SVX RASNIKs.  The possibility of handling these types of effects on a 
single-rasnik basis is discussed in the next section.

\section{Future Improvements}
\label{improvements}

The greatest difficulty in interpreting the RASNIK data from CDF was separating angular tilts 
of the projector piece (which appear as large translations in the raw data) from true 
non-rotational displacements.  The same systems which were 
particularly susceptible to this effect also provided, when examined together, a way to determine 
whether there had in fact been a rotation of the projector piece(s).  It would be nice, however, 
if there were a direct measure of this effect which did not rely on the readout of other 
systems.  One possibility is that future systems might also deploy a small tilt meter on each 
RASNIK projector, or on some select units.  Small tilt meters are commercially available with 
angular resolution of order 1~$\mu$radian.  For the baselines we use, $15-50$~cm, such precision 
would allow corrections to micron precision.  

Another improvement would be to employ a more advanced camera.  During the last few years, the 
state of the art in microvideo image sensors has moved to digital output.  Instead of an NTSC 
output signal, a chip in the same size package is now available which has double the resolution 
and full digital output.  The signal can be transmitted in parallel or clocked out on a single 
cable to a digital input board, removing the need for a frame grabber.  Such a pure digital 
system would also provide a raw data file with the information from each pixel uniquely 
identifiable.  It may even be possible that one could use this information to gauge tilts of the 
projector by examining the `squareness' of a projected square from the coded mask.

\section{Conclusion}
\label{conclusion}

We have developed a modified RASNIK system which is based on two, rather than three elements so 
that ambiguities in motion have been reduced.  We have designed a projector system out of 
radiation-hard and low-mass components so that systems could be deployed in the inner tracker of
the CDF detector.  Radiation damage to the installed RASNIKs has been successfully mitigated by 
operating the systems at a low duty cycle.  The RASNIKs have found large motions due to the 
solenoid and powering of the silicon detectors that would otherwise have been unnoticed since no 
alignment data is taken during those periods.  The RASNIKs have also shown that during 
data-taking periods the entire silicon tracking system is extraordinarily stable on the minutes 
to days time scale, giving us confidence that motions are not degrading CDF's reconstruction or 
triggering ability.  The data-acquisition system has been robust, taking data continuously for 
almost two years with no downtime other than power outages.  The RASNIKs have given CDF a 
minute-by-minute view of motions of its sensitive inner trackers that would otherwise have been 
unavailable.

\section*{Acknowledgments}

We thank Harry van de Graaf for introducing us to the RASNIK system.  We thank Brenna Flaugher, 
Joe Incandela and Rick Snider for advice and support of the RASNIK system within the silicon 
construction project.  We thank Frank Chase for his engineering support during the design and 
construction of prototypes and production pieces.  We are indebted to the technical expertise of 
Greg Derylo, who guided the final installation onto the CDF silicon detectors.  We thank Jim 
Patrick for helping us deploy the DAQ linux box at CDF. This work was supported in part by the 
U.S. Dept. of Energy.

\newpage
\vspace{7.0in}

\newpage
\vspace{7.0in}

\newpage
\vspace{7.0in}

\newpage
\vspace{7.0in}

\newpage
\vspace{7.0in}

\newpage
\vspace{7.0in}

\newpage
\vspace{7.0in}

\newpage
\vspace{7.0in}

\newpage
\vspace{7.0in}

\end{document}